\documentclass[pre,preprintnumbers,twocolumn,showpacs,superscriptaddress]{revtex4}

\usepackage{graphicx}
\usepackage{amsmath}
\usepackage{amssymb}
\usepackage{mathrsfs}
\usepackage[british]{babel}

\begin{document}

\preprint{\today}

\title{Single particle tracking in systems showing anomalous diffusion:
the role of weak ergodicity breaking}

\author{Stas Burov}
\thanks{Submitted to the special issue: Single Molecule Optical studies
of Soft and Complex Matter}
\affiliation{Department of Physics, Bar Ilan University, Ramat-Gan 52900,
Israel}
\author{Jae-Hyung Jeon}
\affiliation{Physics Department, Technical University of Munich, D-85747
Garching, Germany}
\author{Ralf Metzler}
\email{metz@ph.tum.de}
\affiliation{Physics Department, Technical University of Munich, D-85747
Garching, Germany}
\author{Eli Barkai}
\email{barkaie@mail.biu.ac.il}
\affiliation{Department of Physics, Bar Ilan University, Ramat-Gan 52900,
Israel}


\begin{abstract}
Anomalous diffusion has been widely observed by single
particle tracking microscopy in complex systems such as biological cells.
The resulting time series are usually evaluated in terms of time averages.
Often anomalous diffusion is connected with non-ergodic behaviour. In such
cases the time averages remain random variables and hence irreproducible.
Here we present a detailed analysis
of the time averaged mean squared displacement for
systems governed by anomalous diffusion, considering both unconfined and
restricted (corralled) motion. We discuss the behaviour of the time averaged
mean squared displacement for two
prominent stochastic processes, namely, continuous time random walks and
fractional Brownian motion. We also study the distribution of the
time averaged mean squared displacement around its ensemble mean, and show
that this distribution preserves typical process characteristic even for
short time series. Recently, velocity correlation functions were suggested
to distinguish between these processes. We here present analytucal expressions
for the velocity correlation functions. Knowledge of the results presented
here are
expected to be relevant for the correct interpretation of single particle
trajectory data in complex systems.
\end{abstract}

\maketitle

\section{Introduction}

Single particle tracking microscopy provides the position time series
$\mathbf{r}(t)$ of individual particle trajectories in a medium
\cite{braeuchle,saxton,saxton1,qian}.
The information garnered by single particle tracking
yields insights into the mechanisms and forces, that drive or constrain the
motion of the particle. An early example of systematic single particle tracking
is given by the work of Jean Perrin on diffusive motion \cite{perrin}. Due to
the relatively
short individual trajectories, Perrin used an ensemble average over many
trajectories to obtain meaningful statistics. A few years later, Nordlund
conceived a method to record much longer time series \cite{nordlund}, allowing
him to evaluate individual trajectories in terms of the time average and thus
to avoid averages over not perfectly identical particles. Today, single
particle tracking has become a standard tool to characterise the microscopic
rheological properties of a medium \cite{microrheol},
or to probe the active motion of biomolecular motors \cite{motors}. Particularly
in biological cells and complex fluids single particle trajectory methods have
become instrumental in uncovering deviations from normal Brownian motion of
passively moving particles \cite{golding,weber,garini,seisenhuber,elbaum,lene,%
lene1,weiss,weiss1,yves,pan,banks,weihs,weitz}.

\begin{figure}
\includegraphics[width=8.8cm]{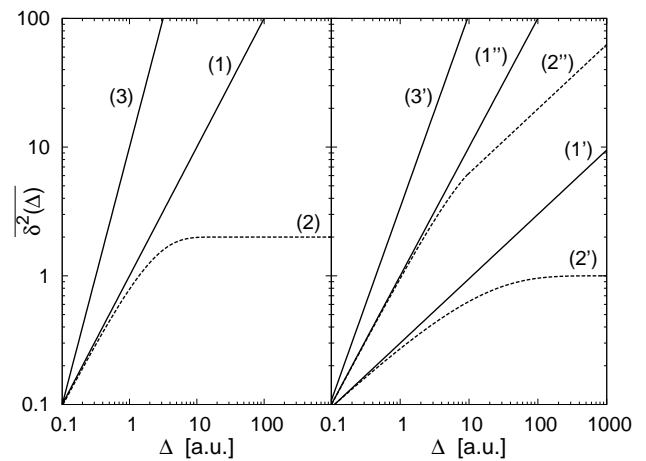}
\caption{Diffusion modes of the time averaged mean squared displacement
(\ref{tamsd}) as function of the lag time $\Delta$. Left: Normal diffusion
growing like $\overline{\delta^2}\simeq\Delta$ (1), restricted
(confined) diffusion with a turnover from $\simeq\Delta$ to $\simeq\Delta^0$
(2), drift diffusion $\simeq\Delta^2$ (3). Right: Ergodic subdiffusion
$\simeq\Delta^{\alpha}$ (1'), restricted ergodic
subdiffusion turning over from $\simeq\Delta^{\alpha}$ to $\simeq\Delta^0$
(2'), non-ergodic subdiffusion $\simeq\Delta$ (1''), restricted non-ergodic
subdiffusion turning over from $\simeq\Delta$ to $\simeq\Delta^{1-\alpha}$
(2''), superdiffusion $\Delta^{1+\alpha}$ (3'). Here, $0<\alpha<1$.
Note the double-logarithmic scale.}
\label{diffusion_modes}
\end{figure}

Classical diffusion patterns are sketched in the left panel of
Fig.~\ref{diffusion_modes}. Accordingly, one may observe free diffusion, leading
to a linear growth with time of the second moment [Line 1 in
Fig.~\ref{diffusion_modes}]. Brownian motion may also
be restricted (corralled, confined). Confinement in a cell, for instance,
could be due to the cell walls. In that case the second moment initially grows
linearly with time and eventually saturates to its thermal value equalling the
second moment of the corresponding Boltzmann distribution [Line 2].
In the presence of a drift the second moment
grows with the square of time [Line 3]. Such
results are typical for simple fluids. In more complex
environments, different patterns may be observed, as displayed on the right of
Fig.~\ref{diffusion_modes}. Here, subdiffusion may occur, for which the
second moment grows slower than linearly with time [Line 1']. Restricted
subdiffusion would depart from this behaviour to reach a plateau [Line 2'].
Driven motion may lead to a superdiffusive power-law form of the second moment
with an exponent between 1 and 2 [Line 3'].  However, as we demonstrate below,
subdiffusion may also be non-ergodic, and the associated time averaged second
moment may grow \emph{linearly\/} with time [Line 1'']. Similarly strange
behaviour may be observed for restricted non-ergodic subdiffusion, which
exhibits a \emph{power-law\/} growth, not a saturation to a plateau [Line 2''].
Non-ergodic processes come along with a significant scatter between
individual trajectories. This is an effect of the ageing nature of the process
that persists for long measurement times.
In the following we discuss in detail the behaviour of passive subdiffusive
motion in terms of time and ensemble averages and address the peculiarities,
that may arise for non-ergodic systems.

\begin{figure}
\includegraphics[width=8.8cm]{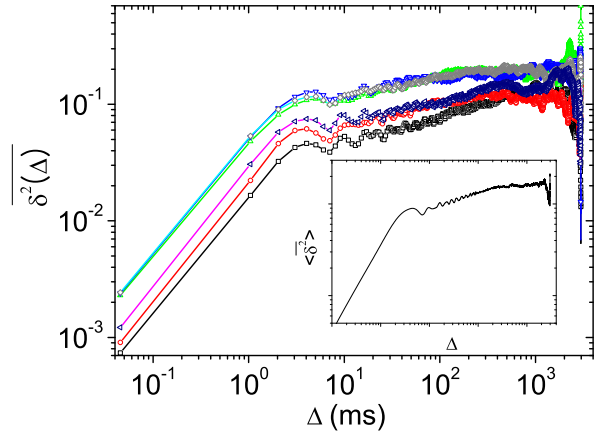}
\caption{Time averaged mean squared displacement of lipid granule motion in
fission yeast cell \emph{S.pombe\/}, measured by optical tweezers.
As function
of the lag time $\Delta$ the initially linear scaling $\overline{\delta
^2}\simeq\Delta$ turns over to the power-law regime $\overline{\delta^2}\simeq
\Delta^{1-\alpha}$, induced by the restoring force exerted on the granules
by the laser trap \cite{christine}. Note also the characteristic scatter
between individual trajectories. Inset: Average behaviour of the shown
trajectories. In both graphs the unit of $\overline{\delta^2(\Delta,T)}$ is
Volt$^2$, i.e., the direct output voltage of the quadrant photodiode. The
voltage is directly proportional to the distance from the centre of the
laser trap.}
\label{christine}
\end{figure}

Non-ergodic behaviour of the above sense is indeed observed experimentally.
Fig.~\ref{christine} shows the time averaged mean squared displacement for
lipid granules in a living fission yeast cell. The motion is recorded by
indirect tracking in an optical tweezers setup. Initially the granule is
located in the bottom of the laser trap potential such that the granule
moves freely. Eventually the granule ventures away from the centre of the
trap and experiences the Hookean trap force. As demonstrated in a detailed
analysis the granule motion indeed exhibits weak ergodicity breaking, giving
rise to the characteristic turnover from an initially linear scaling
$\overline{\delta^2}\simeq\Delta$ with the lag time $\Delta$, to the power-law
regime $\overline{\delta^2}\simeq\Delta^{1-\alpha}$ \cite{christine}.
Moreover a pronounced trajectory-to-trajectory scatter
is observed, again typical for systems with weak ergodicity breaking.

Free diffusion is typically quantified in terms of the second moment.
The mean squared displacement
\begin{equation}
\label{eamsd}
\langle\mathbf{r}^2(t)\rangle\equiv\int\mathbf{r}^2P(\mathbf{r},t)d^3\mathbf{r}
\end{equation}
is obtained as the spatial average over the probability density function
$P(\mathbf{r},t)$ to find the particle at position $\mathbf{r}$ at time $t$.
The quantity
(\ref{eamsd}) therefore corresponds to the ensemble averaged second moment of
the particle position, denoted by angular brackets, $\langle\cdot\rangle$.
In particular, the time $t$ enters into Eq.~(\ref{eamsd}) only as a
parameter. Conversely, single particle trajectories $\mathbf{r}(t)$ are
usually
evaluated in terms of the time averaged mean squared displacement defined as
\begin{equation}
\label{tamsd}
\overline{\delta^2(\Delta,T)}\equiv\frac{1}{T-\Delta}\int_0^{T-\Delta}\Big(
\mathbf{r}(t+\Delta)-\mathbf{r}(t)\Big)^2dt,
\end{equation}
where we use an overline $\overline{\,\,\,\cdot\,\,\,}$ to symbolise the
time average.
Here $\Delta$ is the so-called lag time constituting a time window swept along
the time series, and $T$ is the overall measurement time. The time averaged
mean squared displacement thus compares the particle positions along the
trajectory as separated by the time difference $\Delta$.

In an ergodic system the time average of a certain quantity obtained from
sufficiently long time series is equal to the corresponding ensemble mean
\cite{vankampen,khinchin}. For instance, for the mean squared displacement
ergodicity would imply
\begin{equation}
\lim_{T\to\infty}\overline{\delta^2(\Delta=t,T)}=\langle
\mathbf{r}^2(t)\rangle.
\end{equation}
Brownian motion is ergodic, as well as certain stationary processes leading
to anomalous diffusion, such as fractional Brownian motion considered below.
There exist, however, non-ergodic processes, which
are intimately connected to ageing properties. In what follows we discuss two
prominent models for anomalous diffusion, the stationary fractional Brownian
motion and the ageing continuous time random walk, and analyse in detail the
features of the associated time averaged mean squared displacement.

Generally,
anomalous diffusion denotes deviations from the classical linear dependence
of the mean squared displacement, $\langle\mathbf{r}^2(t)\rangle\simeq t$.
Such anomalies include ultraslow diffusion of the form $\langle\mathbf{r}^2(t)
\rangle\simeq\log^{\beta}t$ \cite{sinai}.
In contrast, anomalous diffusion processes may become faster than ballistic,
for instance for systems with correlated jump lengths or in systems governed
by generalised Langevin equations \cite{tejedor,goychuk}. Here we are
interested in
anomalous diffusion with power-law dependence on time \cite{report},
\begin{equation}
\label{aneamsd}
\langle\mathbf{r}^2(t)\rangle\sim2d\frac{K_{\alpha}}{\Gamma(1+\alpha)}
t^{\alpha},
\end{equation}
for which the anomalous diffusion exponent belongs to the subdiffusive range
$0<\alpha<1$, such that the limit $\alpha=1$ corresponds to Brownian motion.
The proportionality factor $K_{\alpha}$ in Eq.~(\ref{aneamsd}) is the
anomalous diffusion coefficient of physical dimension $\mathrm{cm}^2/\mathrm{
sec}^{\alpha}$. The embedding spatial dimension is $d$, and Eq.~(\ref{aneamsd})
includes the complete Gamma function $\Gamma(z)$.

Subdiffusion of the form (\ref{aneamsd}) with $0<\alpha<1$ occurs in the
following biologically relevant systems.
Fluorescently labelled mRNA in \emph{E.coli\/} bacteria cells was observed
to follow $\overline{\delta^2(\Delta,T)}\simeq\Delta^{\alpha}$ with $\alpha
\approx0.7$ \cite{golding}. This result is consistent with more recent findings
according to which free RNA tracers in living cells exhibit $\alpha\approx0.8$,
while DNA loci show $\alpha=0.4$ \cite{weber}. Telomeres in the nucleus of
mammalian cells were reported to follow anomalous diffusion with $\alpha\approx
0.3$ at shorter times and value $\alpha\approx0.5$ at intermediate times,
before a turnover to normal diffusion occurs
\cite{garini}. Also larger tracer particles show anomalous diffusion, such as
adeno-associated viruses of radius $\approx 15$ nm in a cell with $\alpha=0.5
\ldots0.9$ \cite{seisenhuber} and endogenous lipid granules, of typical size
of few hundred nm with $\alpha\approx0.75\ldots0.85$
\cite{elbaum,lene,lene1}.
It should be noted that this subdiffusion observed by single particle tracking
microscopy is consistent with results from other techniques, such as
fluorescence correlation spectroscopy \cite{banks,weiss,weiss1,yves} or dynamic
light scattering \cite{pan}. Subdiffusion of biopolymers larger than some
10 kD in living cells is due to molecular crowding, the excluded volume effect
in the superdense cellular environment \cite{zimmermann,minton,mcguffee}.
Larger tracer particles also experience subdiffusion due to interaction
with the semiflexible cytoskeleton \cite{weitz}.

Knowledge of the time or ensemble averaged mean squared displacement of
an anomalous diffusion process is insufficient to fully characterise the
underlying stochastic mechanism, as the associated probability density
$P(\mathbf{r},t)$ is no longer necessarily Gaussian, and therefore no longer
specified
by the first and second moments, only \cite{report}. This property is
in contrast to the universal Gaussian nature of Brownian motion which is
effected by the central limit theorem. At the same time the very nature
of the anomalous diffusion process may result in decisively different
behaviours for diffusional mixing and diffusion-limited reactions
\cite{katja}. In
biological cells this would imply significant differences for signalling
and regulatory processes. For a better understanding of the dynamics in
biological cells and other complex fluids knowledge of
the underlying stochastic mechanism is therefore imperative.

Here we discuss the properties of continuous time random walk (CTRW) processes
with diverging characteristic waiting time with respect to their time averaged
behaviour, expanding on our earlier work \cite{he,pnas,appb}. We show that for
free motion the lag time dependence of the time averaged mean squared
displacement, $\overline{\delta^2(\Delta,T)}\simeq\Delta$, is insensitive to
the anomalous diffusion exponent $\alpha$ for CTRW processes. In contrast,
for confined CTRW subdiffusion a universal scaling behaviour emerges,
$\overline{\delta^2(\Delta,T)}\simeq\Delta^{1-\alpha}$, with dynamic exponent
$1-\alpha$ [curves (1'') and (2'') in
Fig.~\ref{diffusion_modes}]. Subdiffusion governed by fractional Brownian
motion (FBM) leads to the scaling $\overline{\delta^2(\Delta,T)}\simeq\Delta^{
\alpha}$ of the time averaged mean squared displacement, turning over to a
saturation plateau under confinement, $\overline{\delta^2(\Delta,T)}\simeq
\Delta^0$ [curves (1') and (2') in Fig.~\ref{diffusion_modes}].

We particularly emphasise the irreproducible nature of the time averaged
quantities and their associated scatter around the ensemble mean for CTRW
subdiffusion processes. This
randomness of the time averages is captured by the distribution function
of the amplitude of the time average. We show that even for relatively short
trajectories this distribution is a good characteristic for the underlying
process. In contrast for FBM processes the scatter typical of many
single particle experiments is not found in the long time limit.

In the remainder of this work, for simplicity we restrict the discussion to
the one-dimensional case ($d=1$). Generalisation to higher dimensions is
straightforward.
The article is structured as follows. We start with a brief introduction to
CTRW and FBM. We then consider the cases of unbounded motion and confined
anomalous diffusion in the subsequent two Sections. The distribution of the
time averages will be presented thereafter. Finally, we discuss the velocity
autocorrelation functions for subdiffusive CTRW and FBM processes, before
presenting a concluding discussion.

\section{Anomalous diffusion processes}

Although both CTRW and FBM give rise to an ensemble averaged mean squared
displacement of the form (\ref{eamsd}), they are fundamentally different
processes, as outlined here. We note in passing that also in the random
motion on a fractal support subdiffusion arises \cite{havlin,kimmich}.
We will not pursue this type of anomalous diffusion in the following.

\subsection{Continuous time random walk}

CTRW theory dates back to the work of Montroll and Weiss \cite{montroll},
and was championed in the analysis of charge carrier motion in amorphous
semiconductors by Scher and Montroll \cite{scher}.
CTRW has become a standard statistical tool
to describe processes ranging from particle motion in actin networks
\cite{weitz} to the tracer motion in groundwater \cite{scher1}.

Each jump of a CTRW process is characterised by a random jump length and a
random waiting time elapsing before the subsequent jump. At each jump the
jump length and waiting time are chosen independently. For a subdiffusive
process we assume that the variance of the jump lengths is given by a
finite value $\langle\delta x^2\rangle$, and we consider the unbiased
case $\langle\delta x\rangle=0$. On a lattice of spacing $a$, we would
have $\langle\delta x^2\rangle=a^2$. In contrast, the waiting times
$\tau$ are drawn from the probability density
\begin{equation}
\label{wtd}
\psi(\tau)\simeq\frac{\overline{\tau}^{\alpha}}{|\Gamma(-\alpha)|}\tau^{-1
-\alpha}
\end{equation}
for large $\tau$,
with $0<\alpha<1$. This form of $\psi$ is scale-free, that is, the average
waiting time $\langle\tau\rangle$ diverges, causing effects such as ageing
\cite{rel,ageing} and weak ergodicity breaking \cite{web}. In Eq.~(\ref{wtd}),
the quantity $\overline{\tau}$ is a scaling factor. The anomalous diffusion
constant in this case becomes \cite{report,mebakla}
\begin{equation}
\label{an_coff}
K_{\alpha}=\frac{\langle\delta x^2\rangle}{2\overline{\tau}^{\alpha}}.
\end{equation}

\begin{figure}
\includegraphics[width=8.8cm]{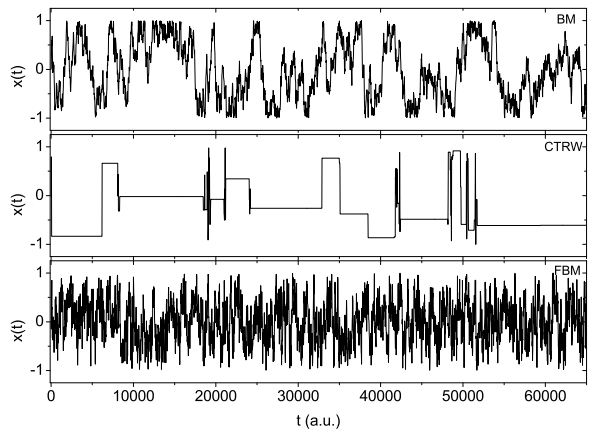}
\caption{Time series $x(t)$ for Brownian motion (top), CTRW (middle), and
FBM (bottom). The anomalous diffusion exponent for FBM and CTRW is $\alpha
=0.5$. For CTRW the stalling events are outstanding, while in the case of
FBM strong antipersistence occurs. Note that the stalling events and the
antipersistence are less pronounced for larger values of $\alpha$.
Typical experimental data include additional noise, such that the
appearance of measured data will not display the ideal behaviour shown here.}
\label{timeseries}
\end{figure}

The scale-freeness of $\psi(\tau)$ allows individual waiting times $\tau$ to
become quite large. No matter how long a given time series is chosen, single
$\tau$ values may become of the order of the length of the entire time span
covered in the trajectory. An example is shown in Fig.~\ref{timeseries}:
the stalling events with large waiting times $\tau$ are quite distinct.
For values of the anomalous diffusion exponent $\alpha$ that are closer to 1
the stalling is less pronounced.
Physically, the power-law form of the waiting time distribution $\psi(\tau)$
may be related to comb models \cite{havlin} or random energy landscapes
\cite{rel,rel1}; for more details, compare Refs.~\cite{hughes,bouchaud,scher2,%
klages}.

\subsection{Fractional Brownian motion}

FBM is a Gaussian process with stationary increments. Its position is
defined in terms of the Langevin equation
\begin{equation}
\frac{dx(t)}{dt}=\xi(t),
\end{equation}
or, alternatively,
\begin{equation}
x(t)=\int_0^t\xi(t')dt'.
\end{equation}
The motion is driven by stationary, fractional Gaussian noise $\xi(t)$ with
zero mean $\langle\xi(t)\rangle$ and long-ranged noise correlation \cite{fbm}
\begin{eqnarray}
\nonumber
\langle\xi(t_1)\xi(t_2)\rangle&=&\alpha K_{\alpha}^*(\alpha-1)|t_1-t_2|^{\alpha
-2}\\
&&+2\alpha K_{\alpha}^*|t_1-t_2|^{\alpha-1}\delta(t_1-t_2),
\label{fbnoise}
\end{eqnarray}
contrasting the uncorrelated noise for normal diffusion $\alpha=1$: $\langle
\xi(t_1)\xi(t_2)\rangle=2K_1\delta(t_1-t_2)$.
In Eq.~(\ref{fbnoise}) the anomalous diffusion
exponent is connected to the traditionally used
Hurst exponent by $H=\alpha/2$, and we introduce
the abbreviation $K_{\alpha}^*=K_{\alpha}/\Gamma(1+\alpha)$ for consistency
with standard FBM notation. For subdiffusion the fractional Gaussian noise is
anticorrelated, decaying like $\langle\xi(t_1)\xi(t_2)\rangle\sim-K_{\alpha}^*
\alpha|\alpha-1||t_1-t_2|^{\alpha-2}$. This implies that a given step is likely
to go into the direction opposite to the previous step. The corresponding
oscillating behaviour is seen in Fig.~\ref{timeseries}.
The position autocorrelation function of FBM becomes
\begin{equation}
\label{fbm_auto}
\langle x(t_1)x(t_2)\rangle=K_{\alpha}^*\Big(t_1^{\alpha}+t_2^{\alpha}-|t_1
-t_2|^{\alpha}\Big),
\end{equation}
so that at equal times $t_1=t_2$ we recover the mean squared displacement
(\ref{aneamsd}).


If the fractional Gaussian noise is not considered external, but the validity
of the fluctuation-dissipation theorem is imposed, one obtains the generalised
Langevin equation (GLE) \cite{kubo},
\begin{equation}
\label{gle}
m\frac{d^2x(t)}{dt^2}=-\overline{\gamma}\int_0^t(t-t')^{\beta-2}\frac{dx}{dt'}
dt'+\eta\xi(t),
\end{equation}
where we write $\beta$ instead of the exponent $\alpha$ in Eq.~(\ref{fbnoise}).
In the GLE (\ref{gle}), we defined the coupling coefficient $\eta=\sqrt{k_B
\mathcal{T}\overline{\gamma}/[\beta K_{\beta}^*(\beta-1)]}$ according to the
fluctuation-dissipation theorem, where $k_B$ is the Boltzmann constant and
$\mathcal{T}$ the absolute temperature. In what follows we only consider
the overdamped limit, in which the inertia term $md^2x(t)/dt^2$ can be
neglected. The GLE then gives rise to the form
\begin{equation}
\langle x^2(t)\rangle\simeq t^{2-\beta}
\end{equation}
of the mean squared displacement. In contrast to FBM, that is, the GLE leads
to subdiffusion for persistent noise with $1<\beta<2$, while $0<\beta<1$
yields superdiffusion.

FBM and the related GLE are used to describe processes such as long term
storage capacity of water reservoirs \cite{hurst}, climate fluctuations
\cite{climate}, economical market dynamics \cite{eco}, single file diffusion
\cite{singlefile}, and elastic models \cite{chechkin}. Motion of this type
has also been associated with the relative motion of aminoacids in proteins
\cite{xie}, and the free diffusion of biopolymers under molecular crowding
conditions \cite{weiss1,weber,marcin}.

\section{Free anomalous diffusion}

Let us begin with considering anomalous diffusion on an infinite domain and
without drift. To find an analytical expression for the time averaged mean
squared displacement (\ref{tamsd}) we note that even a Brownian process
recorded over a finite time span $T$ will show fluctuations in the number
of jumps performed during $T$. To average out these trajectory-to-trajectory
fluctuations we introduce the ensemble mean,
\begin{equation}
\label{eatamsd}
\left<\overline{\delta^2(\Delta,T)}\right>=\frac{1}{T-\Delta}\int_0^{T-
\Delta}\left<\Big(x(t+\Delta)-x(t)\Big)^2\right>dt.
\end{equation}
We can then express the integrand in terms of the variance of the jump
lengths, $\langle\delta x^2\rangle$, and the average number of jumps
$n(t,t+\Delta)$ in the time interval $(t,t+\Delta)$, as follows:
\begin{equation}
\left<\Big(x(t+\Delta)-x(t)\Big)^2\right>=\langle\delta x^2\rangle
n(t,t+\Delta).
\end{equation}
For a regular random walk on average every jump occurs after the waiting time
$\langle\tau\rangle$. Thus $n(t,t+\Delta)=\Delta/\langle\tau\rangle$, and
\begin{equation}
\label{btamsd}
\left<\overline{\delta^2(\Delta,T)}\right>=2K_1\Delta,
\end{equation}
where we defined the diffusion constant $K_1=\langle\delta x^2\rangle/[2\langle
\tau\rangle]$.
In the Brownian limit the time averaged mean squared displacement (\ref{btamsd})
in terms of the lag time $\Delta$ takes on exactly the same form as the ensemble
averaged mean squared displacement (\ref{eamsd}) as function of time $t$. This
is not surprising, as Brownian motion is an ergodic process. For sufficient
duration $T$ of the time records any time average converges to the corresponding
ensemble average, and thus the ensemble average in expression (\ref{eatamsd})
is no longer necessary.

\subsection{Continuous time random walk}

\begin{figure}
\includegraphics[width=8.8cm]{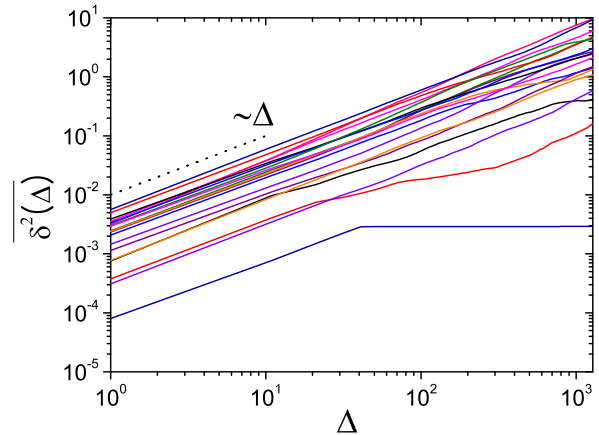}
\caption{Time averaged mean squared displacement for unconfined CTRW motion
with $\alpha=0.5$, shown for 20 individual trajectories. The overall
measurement time is $T=10^5$ (a.u.). Note the local changes of the slope in the
trajectories as well as the complete stalling in the lowest curve, all
bearing witness to the scale-free nature of the underlying waiting time
distribution $\psi(\tau)\simeq\overline{\tau}^{\alpha}/\tau^{1+\alpha}$.}
\label{ctrw_tamsd}
\end{figure}

The number of jumps of a CTRW process with waiting time distribution of the form
(\ref{wtd}) on average grows sublinearly with time, $n(0,t)\sim t^{\alpha}/[
\overline{\tau}^{\alpha}\Gamma(1+\alpha)]$ \cite{hughes}. This time evolution
translates
into the mean squared displacement (\ref{eamsd}) with the anomalous diffusion
coefficient (\ref{an_coff}).
Combining the time dependence of $n(0,t)$ for CTRW subdiffusion with the
definition (\ref{eatamsd}) we obtain the following result for the time averaged
mean squared displacement,
\begin{equation}
\label{ctrwtamsd}
\left<\overline{\delta^2(\Delta,T)}\right>\sim2K_{\alpha}\frac{\Delta}{T^{1-
\alpha}}
\end{equation}
in the limit $\Delta\ll T$ \cite{he,lubelski}. This follows from expansion of
the relation $n(t,t+\Delta)=n(0,t+\Delta)-n(0,t)$. The noteworthy feature of
this result is that the linear lag time dependence of normal diffusion remains
completely unaffected by the anomalous nature of the stochastic process. Only
the dependence on the overall measurement time $T$ witnesses the underlying
subdiffusion. Eq.~(\ref{ctrwtamsd}) can be understood if we notice that for
normal diffusion we have $\delta^2(\Delta,T)\simeq\Delta/\langle\tau\rangle$
[since according to Einstein $K_1\propto1/\langle\tau\rangle$] where $\langle
\tau\rangle$ is the mean time between jump events. Now, for the subdiffusive
case $\langle\tau\rangle$ diverges and must be replaced by $\int_0^T\psi(\tau)
\tau d\tau\propto T^{1-\alpha}$, which explains the term $\delta^2\propto
\Delta/T^{1-\alpha}$. Fig.~\ref{ctrw_tamsd} shows 20 trajectories of a CTRW
process with $\alpha=0.5$, displaying the general trend $\overline{\delta^2
(\Delta,T)}\simeq\Delta$.

Rewriting the result (\ref{ctrwtamsd}) in the form
\begin{equation}
\label{rew}
\left<\overline{\delta^2(\Delta,T)}\right>\sim2\overline{K}(T)\Delta
\end{equation}
we see that the effective diffusion coefficient $\overline{K}(T)$ decays as
function of the measurement time $T$. The longer the system evolves after
its initial preparation, the less mobile it appears, consistent with the
ageing property of CTRW subdiffusion. For instance, in the picture of the
trapping events, in the course of time deeper and deeper traps may be
encountered by the diffusing particle, such that it gets more and more
stuck. This increasing immobility is mirrored in the form $\overline{K}(T)$.
Note that Eqs.~(\ref{ctrwtamsd}) and (\ref{rew}) suggest
that from measuring the $\Delta$
dependence of an anomalous diffusion process one might draw the erroneous
conclusion that normal diffusion were observed.

The disparity between the time averaged mean squared displacement
(\ref{ctrwtamsd}) and its ensemble averaged counterpart (\ref{aneamsd})
demonstrates the weak ergodicity breaking characteristic of a process with
diverging time scale \cite{rel,web}. Single or few, long waiting times are
also responsible for the pronounced deviations between individual realisations
shown in Fig.~\ref{ctrw_tamsd}. This apparent irreproducibility of the time
averaged mean squared displacement is again intimately coupled to the weak
ergodicity breaking nature of the CTRW subdiffusion process. We will discuss
this feature more quantitatively in terms of the distribution $\phi_{\alpha}
(\xi)$ as function of the relative deviation $\xi=\overline{\delta^2}/\left<
\overline{\delta^2}\right>$ of the time averaged mean squared displacement
around its ensemble mean in Section \ref{scatter}.


\subsection{Fractional Brownian motion}

\begin{figure}
\includegraphics[width=8.8cm]{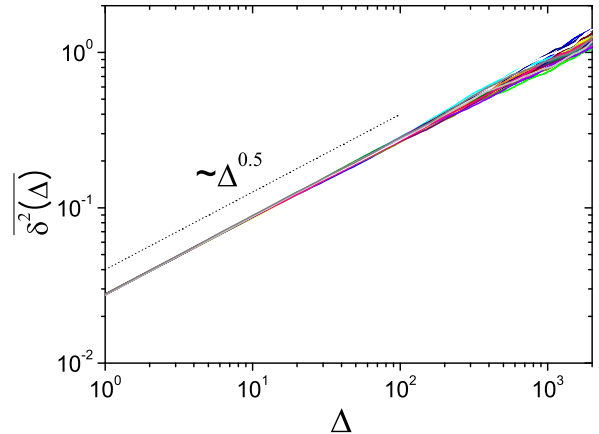}
\caption{Time averaged mean squared displacement for unconfined FBM with
$\alpha=0.5$. The scatter between individual trajectories is minute,
mirroring the ergodicity of this process.}
\label{fbm_tamsd}
\end{figure}

In contrast to the above behaviour of CTRW subdiffusion processes, FBM is
ergodic. Indeed, by help of the position autocorrelation (\ref{fbm_auto})
the time averaged mean squared displacement (\ref{eatamsd}) becomes \cite{deng}
\begin{equation}
\label{fbm_free}
\left<\overline{\delta^2(\Delta,T)}\right>=2K_{\alpha}^*\Delta^{\alpha},
\end{equation}
exactly matching the ensemble averaged mean squared displacement, $\langle
x^2(t)\rangle=2K_{\alpha}^*t^{\alpha}$. However,
ergodicity is reached algebraically slowly \cite{deng}, see also below. In
Fig.~\ref{fbm_tamsd} the comparatively minute scatter between individual
trajectories supports the ergodic behaviour of FBM processes. The somewhat
larger deviations at longer lag times are due to worsening statistics when
$\Delta\to T$.

\section{Confined anomalous diffusion}

Since the motion of tracer particles is typically confined, for example, by
the cell walls or internal membranes, we now consider the important case of
anomalous diffusion in a bounded domain.

In a finite interval $[-L,L]$ the mean squared displacement of a Brownian
particle initially released well away from the boundaries will grow
linearly in time and eventually turn over to a stationary plateau of
magnitude $\langle x^2\rangle_{\mathrm{st}}=L^2/3$. Similarly, if the
particle evolves in the confinement of an external harmonic potential
of the form $V(x)=\frac{1}{2}m\omega^2x^2$, the thermal value $\langle x^2
\rangle_{\mathrm{th}}=k_B\mathcal{T}/[m\omega^2]$ will eventually be attained.

\subsection{Continuous time random walk}

How is this behaviour modified for a CTRW subdiffusion process? Under the
influence of an arbitrary external potential $V(x)=-\int^x F(x')dx'$ defining
the force $F(x)$, CTRW
subdiffusion can be described by the fractional Fokker-Planck equation
\cite{mebakla,report}, or, equivalently, in terms of the following coupled
Langevin equations \cite{subord}:
\begin{subequations}
\begin{eqnarray}
\label{le1}
\frac{dx(s)}{ds}&=&\frac{K}{k_B\mathcal{T}}F(x)+\eta(s)\\
\frac{dt(s)}{ds}&=&\omega(s).
\label{le2}
\end{eqnarray}
\end{subequations}
Here the position $x$ is expressed in terms of the parameter $s$ (the internal
time), and driven by the white Gaussian noise $\eta(s)$. Thus, Eq.~(\ref{le1})
defines standard Brownian motion $x(s)$, where $K$ is the diffusivity for the
normal diffusion process in internal time $s$. Laboratory time $t$ is
introduced by the so-called subordination through the process $\omega(s)$,
given by the probability density function \cite{report}
\begin{equation}
p_t(s)=\frac{1}{\alpha}\left(\frac{K_{\alpha}}{K}\right)^{1/\alpha}\frac{t}{
s^{1+1/\alpha}}l_{\alpha}\left(\left[\frac{K_{\alpha}}{K}\right]^{1/\alpha}
\frac{t}{s^{1+1/\alpha}}\right),
\end{equation}
where $l_{\alpha}(z)$ is a one-sided L{\'e}vy stable probability density
with Laplace transform $\int_0^{\infty}l_{\alpha}(z)\exp(-uz)dz=\exp(-u^
{\alpha})$. Thus, Eq.~(\ref{le2}) transforms the Brownian process $x(s)$
with diffusivity $K$ into the subdiffusive motion with generalised diffusivity
$K_{\alpha}$. On the basis
of this scheme, the ensemble averaged position-position correlation in
an arbitrary confining potential $V(x)$ becomes
\begin{equation}
\label{ctrw_conf_ea_auto}
\langle x(t_1)x(t_2)\rangle=\Big(\langle x^2\rangle_B-\langle x\rangle_B^2
\Big)\frac{B(t_1/t_2,\alpha,1-\alpha)}{\Gamma(\alpha)\Gamma(1-\alpha)}+
\langle x\rangle_B^2,
\end{equation}
valid in the limit $t_2\ge t_1\gg(1/[K_{\alpha}\lambda_1])^{1/\alpha}$,
where $\lambda_1$ is the smallest non-zero eigenvalue of the corresponding
Fokker-Planck operator \cite{pnas}. Eq.~(\ref{ctrw_conf_ea_auto}) demonstrates
that, despite the confinement, the process is non-stationary.
In result (\ref{ctrw_conf_ea_auto}) we used the incomplete Beta function
\begin{equation}
B(t_1/t_2,\alpha,1-\alpha)\equiv\int_0^{t_1/t_2}z^{\alpha-1}(1-z)^{-\alpha}
dz
\end{equation}
and defined the first and second Boltzmann moments, whose general definition
is
\begin{equation}
\langle x^n\rangle_B=\frac{1}{\mathscr{Z}}\int_{-\infty}^{\infty}x^n\exp\left(
-\frac{V(x)}{k_B\mathcal{T}}\right)dx.
\end{equation}
The partition function reads
\begin{equation}
\mathscr{Z}=\int_{-\infty}^{\infty}\exp\left(-\frac{V(x)}{k_B\mathcal{T}}
\right)dx.
\end{equation}
At large time separation, $t_2\gg t_1$, the position autocorrelation decays
algebraically,
\begin{eqnarray}
\nonumber
\langle x(t_1)x(t_2)\rangle&\sim&\\
&&\hspace*{-2.0cm}\Big(\langle x^2\rangle_B-\langle x\rangle_B^2
\Big)\frac{(t_1/t_2)^{\alpha}}{\alpha\Gamma(1+\alpha)\Gamma(1-\alpha)}+
\langle x\rangle_B^2,
\end{eqnarray}
towards the value $\langle x\rangle_B^2$.
The corresponding limiting behaviour of the incomplete Beta
function reads
\begin{equation}
\label{bet_lim}
\frac{B(t/(t+\Delta),\alpha,1-\alpha)}{\Gamma(\alpha)\Gamma(1-\alpha)}\sim1-
\frac{\sin(\pi\alpha)}{(1-\alpha)\pi}\left(\frac{\Delta}{t}\right)^{1-\alpha}.
\end{equation}

Inserting expression (\ref{ctrw_conf_ea_auto}) into the definition of the time
averaged mean squared displacement (\ref{eatamsd}) we obtain \cite{pnas}
\begin{eqnarray}
\nonumber
\left<\overline{\delta^2(\Delta,T)}\right>&=&\frac{1}{T-\Delta}\int_0^{T-\Delta}
\Big[\langle x(t+\Delta)x(t+\Delta)\rangle\\
&&\hspace*{-0.8cm}+\langle x(t)x(t)\rangle-2\langle x(t+\Delta)x(t)\rangle
\Big]dt.
\end{eqnarray}
Then, with relation (\ref{bet_lim}) we arrive at the scaling behaviour
\begin{equation}
\label{conf_tamsd}
\left<\overline{\delta^2(\Delta,T)}\right>\sim\Big(\langle x^2\rangle_B-
\langle x\rangle_B^2\Big)\frac{2\sin(\pi\alpha)}{(1-\alpha)\pi\alpha}\left(
\frac{\Delta}{T}\right)^{1-\alpha},
\end{equation}
valid in the limits $\Delta/T\ll1$ and $\Delta\gg(1/[K_{\alpha}\lambda_1])
^{1/\alpha}$.
Result (\ref{conf_tamsd}) is quite remarkable: instead of the naively assumed
saturation toward the stationary plateau
a power-law growth $\left<\overline{\delta^2(\Delta,T)}\right>\sim(\Delta/T)^{
1-\alpha}$ is observed. Only when the lag time $\Delta$ approaches the overall
measurement time $T$ the singularity in expression (\ref{eatamsd}) causes a
dip toward the plateau of the ensemble averaged mean squared displacement.
Additionally Eqs.~(\ref{ctrw_conf_ea_auto}) and (\ref{conf_tamsd}) are
universal in the sense that the exact form of the confining potential solely
enters into the prefactor through the first two moments of the Boltzmann
distribution corresponding to the confining potential $V(x)$.
We note that the asymptotic scaling $\simeq(\Delta/T)^{1-\alpha}$ is
consistent with the numerical analysis presented in Ref.~\cite{igor}.

\begin{figure}
\includegraphics[width=8.8cm]{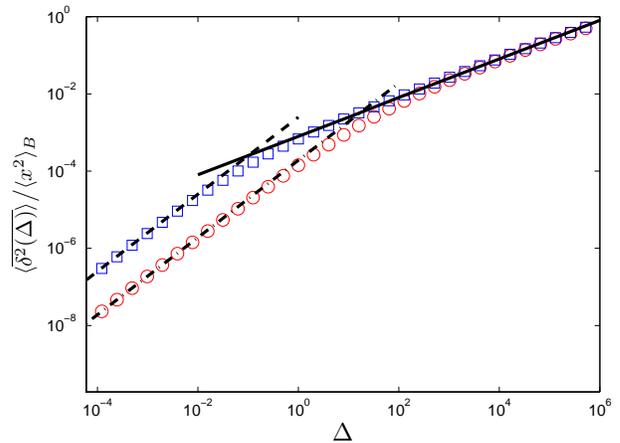}
\caption{Simulated behaviour of $\left<\overline{\delta^2(\Delta)}\right>$ for
an harmonic binding potential $V(x)= 2x^2$ ($\square$) and a particle in a box
of length $2$ ($\bigcirc$), with $\alpha=1/2$, measurement time $T=10^7$
(a.u.), $k_B
\mathcal{T}=0.1$, and $K_{0.5}=0.0892$. Without fitting, the lines show the
analytic results for the transition from $\simeq\Delta^1$ for short lag times
according to
Eq.~(\ref{ctrwtamsd}) ($--$ and $-\cdot-$) to $\simeq\Delta^{1-\alpha}$ for
long lag times, Eq.~(\ref{conf_tamsd}) (---). For long $\Delta$, $\left<
\overline{\delta^2(\Delta)}\right>/\langle x^2\rangle_B$ exhibits universal
behaviour, in the sense that the curve does not depend on the external field.}
\label{confined_ctrw}
\end{figure}

Fig.~\ref{confined_ctrw} depicts the ensemble mean of the time averaged mean
squared displacement (\ref{eatamsd}) for two types of confinement, an harmonic
potential and a box potential. The particle is initially placed at the bottom
of the potential well and in the middle of the box potential, respectively.
This is why at short lag times the particle exhibits the linear scaling
$\left<\overline{\delta^2(\Delta,T)}\right>\simeq\Delta$ typical for
unconfined motion, before turning over to the confinement-induced scaling
$\simeq\Delta^{1-\alpha}$. Note that this plot is fit-free, i.e., the
theoretically calculated asymptotic behaviours nicely fall on top of the
simulations results.

\begin{figure}
\includegraphics[width=8.8cm]{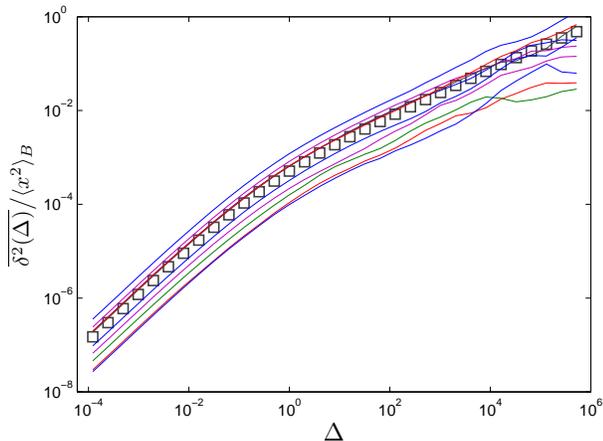}
\caption{Scatter between individual trajectories in the harmonic potential
$V(x)=2x^2$. The extent of the deviations between the trajectories does not
change qualitatively between the initial unbiased motion and the later
confinement-dominated regime. The squares ($\square$) represent simulations
results for the ensemble average $\left<\overline{\delta^2(\Delta,T)}\right>$.
Same parameters as in Fig.~\ref{confined_ctrw}.}
\label{conf_scatter}
\end{figure}

As shown in Fig.~\ref{conf_scatter} individual trajectories still exhibit the
pronounced scatter typical for CTRW subdiffusion. Visually the scatter does
not change between the unbiased initial motion and the confinement-dominated
part of the process. This is due to the fact that the scatter is caused by
the scale-freeness of the waiting times. In our process waiting times and
jump lengths are decoupled, corresponding to the subordination property of
CTRW subdiffusion \cite{report}.

\subsection{Continuous time random walk with waiting time cutoff}

\begin{figure}
\includegraphics[width=8.8cm]{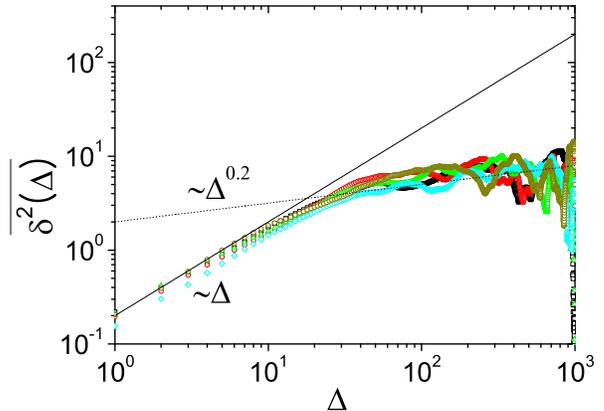}
\caption{CTRW subdiffusion for a waiting time with exponential cutoff,
Eq.~(\ref{wtd_cutoff}), in the box $[-3,3]$ with reflecting boundary
conditions. The generic
behaviour of initial and final scaling, $\overline{\delta^2(\Delta,T)}\simeq
\Delta$ and $\simeq\Delta^{1-\alpha}$ remains unaltered. We chose $\alpha=0.8$,
$\overline{\tau}=1$, $\tau^*=20$ and overall measurement time $T=1000$
(a.u.).}
\label{ctrw_cutoff_conf}
\end{figure}

What happens if we introduce a cutoff in the power-law of the waiting time
distribution of the form
\begin{equation}
\label{wtd_cutoff}
\psi(\tau)=\frac{d}{d\tau}\left[1-\frac{\overline{\tau}^{\alpha}}{(
\overline{\tau}+\tau)^{\alpha}}\exp\Big(-\tau/\tau^*\Big)\right],
\end{equation}
in which a characteristic time scale $\tau^*$ is introduced, terminating
the power-law scaling? For the waiting time density (\ref{wtd_cutoff}) the
characteristic waiting time $\int_0^{\infty}\tau\psi(\tau)d\tau$ becomes
finite.
At sufficiently short times one would expect this process to still exhibit
the features of CTRW subdiffusion, while at times $\tau\gg\tau^*$ the
process should converge to regular Brownian motion with a Gaussian propagator.
In Fig.~\ref{ctrw_cutoff_conf} we demonstrate that for a suitable choice of
the cutoff time $\tau^*$ the behaviour of the time averaged mean squared
displacement $\overline{\delta^2(\Delta,T)}$ under confinement preserves the
characteristic non-ergodic features of CTRW subdiffusion, i.e., the turnover
from the
initial scaling $\simeq\Delta$ to the confinement-dominated scaling $\Delta^{
1-\alpha}$. Below we will show that at the same time the distribution of the
time average around its ensemble mean is significantly altered.

\subsection{Fractional Brownian motion}

Ergodicity remains unaffected for FBM, that is confined to an interval of size
$[-L,L]$. Namely, convergence of $\overline{\delta^2(\Delta,T)}$ is observed
toward the stationary value
$\langle x^2\rangle_{\mathrm{st}}=L^2/3$ \cite{jae}. For FBM under the
influence of an harmonic potential $V(x)=\frac{1}{2}m\omega^2x^2$ the
position-position correlation can be calculated exactly \cite{oleksii}.
Inserting into the ensemble mean (\ref{eatamsd}) of the time averaged
mean squared displacement one can show that the initial behaviour
$\langle x^2(t)\rangle\simeq t^{\alpha}$ turns over to
the stationary value $\langle x^2\rangle_{\mathrm{st}}=\Gamma(\alpha+1)
k_B\mathcal{T}/[m\omega^2]$. In general, the time averaged mean squared
displacement for confined FBM converges to a constant:
\begin{equation}
\label{fbm_cnf}
\left<\overline{\delta^2(\Delta,T)}\right>\sim\mathrm{const.}
\end{equation}

\section{Fluctuations of the time average and ergodicity breaking parameter}
\label{scatter}

The deviations of the time averaged mean squared displacement $\overline{\delta
^2(\Delta,T)}$ between individual trajectories can be quantified in terms of
the probability density function $\phi_{\alpha}(\xi)$ of the dimensionless
ratio
\begin{equation}
\xi\equiv\frac{\overline{\delta^2(\Delta,T)}}{\left<\overline{\delta^2(\Delta,
T)}\right>}
\end{equation}
of the time averaged mean squared displacement over its ensemble mean. For
an ergodic process this distribution is necessarily sharp,
\begin{equation}
\label{phi1}
\phi_{\mathrm{erg}}(\xi)=\delta(\xi-1),
\end{equation}
for long measurement time $T$. Deviations from this form are expected for
non-ergodic processes such as CTRW subdiffusion, but also for relatively
short trajectories. We here address both effects and demonstrate that the
distribution $\phi_{\alpha}(\xi)$ is a quite reliable means to distinguish
different stochastic processes even when the recorded time series are fairly
short.

\subsection{Continuous time random walk}

For CTRW subdiffusion with power-law waiting time distribution (\ref{wtd}) the
distribution assumes the form \cite{he,igor_epl}
\begin{equation}
\phi_{\alpha}(\xi)=\frac{\Gamma(1+\alpha)^{1/\alpha}}{\alpha\xi^{1+1/\alpha}}
l_{\alpha}\left(\frac{\Gamma(1+\alpha)^{1/\alpha}}{\xi^{1/\alpha}}\right)
\end{equation}
for $T\to\infty$,
where $l_{\alpha}(z)$ is again a one-sided L{\'e}vy stable distribution, whose
Laplace transform is $\mathscr{L}\{l_{\alpha}(z)\}=\exp\left(-u^{\alpha}
\right)$. Special cases include the Brownian limit (\ref{phi1}) for $\alpha
=1$ and the Gaussian shape
\begin{equation}
\label{phi_gauss}
\phi_{1/2}(\xi)=\frac{2}{\pi}\exp\left(-\frac{\xi^2}{\pi}\right)
\end{equation}
for $\alpha=1/2$.
In fact, we observe an exponentially fast decay of $\phi_{\alpha}(\xi)$ with
large $\xi$ for all $0<\alpha<1$, compare Appendix \ref{app1}.

\begin{figure}
\includegraphics[width=8.8cm]{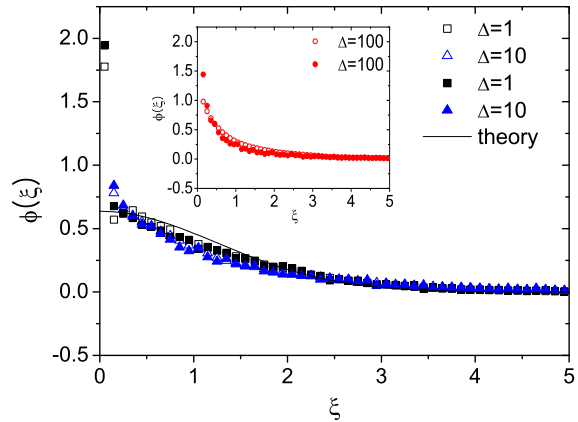}
\caption{Distribution $\phi_{\alpha}(\xi)$ of the time averaged mean squared
displacement, with $\xi=\overline{\delta^2}/\left<\overline{\delta^2}\right>$
for a CTRW process with $\alpha=0.5$, $\overline{\tau}=1$, and $T=128$ (a.u.).
The filled and open symbols, respectively, represent the unconfined case and
confined motion on the interval $[-2,2]$. We see quite good agreement with the
expected Gaussian limiting distribution of the scatter, Eq.~(\ref{phi_gauss}),
centred around $\xi=0$. In the inset we show the case for the largest measured
lag time, $\Delta=100$, also in good agreement with the predicted shape of the
distribution.}
\label{ctrw_scatter}
\end{figure}

For short trajectories the basic shape of the distribution $\phi_{\alpha}(\xi)$
is surprisingly well preserved \cite{jae_scatter}. This is demonstrated in
Fig.~\ref{ctrw_scatter}, in which the characteristic asymmetric shape with
respect to the ergodic value $\xi=1$ for the case $\alpha=0.5$ is reproduced
for a process with $T=128$, even for a lag time as long as $\Delta
=100$. No significant dependence on the confinement is observed. Also for larger
values of $\alpha$ the finite value for small $\xi$ is similarly reproduced
for short trajectories \cite{jae_scatter}, pointing at the quite remarkably
reliability on the shape of the scatter distribution $\phi_{\alpha}(\xi)$
for CTRW subdiffusion. As we show below, the distribution for FBM processes
with its zero value at $\xi=0$ can be clearly distinguished from the CTRW
form.

A useful parameter to quantify the violation of ergodicity is the ergodicity
breaking parameter \cite{he}
\begin{equation}
\label{eb}
\mathrm{EB}=\lim_{T\to\infty}\frac{\left<\left(\overline{\delta^2}\right)^2
\right>-\left<\overline{\delta^2}\right>^2}{\left<\overline{\delta^2}\right>
^2}=\frac{2\Gamma(1+\alpha)^2}{\Gamma(1+2\alpha)}-1
\end{equation}
EB varies from $\mathrm{EB}=1$ for $\alpha\to0$ monotonically to $\mathrm{EB}=
0$ in the Brownian limit $\alpha=1$. For the special case $\alpha=1/2$ one
finds $\mathrm{EB}=\pi/2-1\approx0.57$, while for $\alpha=0.75$, $\mathrm{EB}
\approx0.27$.

\subsection{Continuous time random walk with waiting time cutoff}

\begin{figure}
\includegraphics[width=8.8cm]{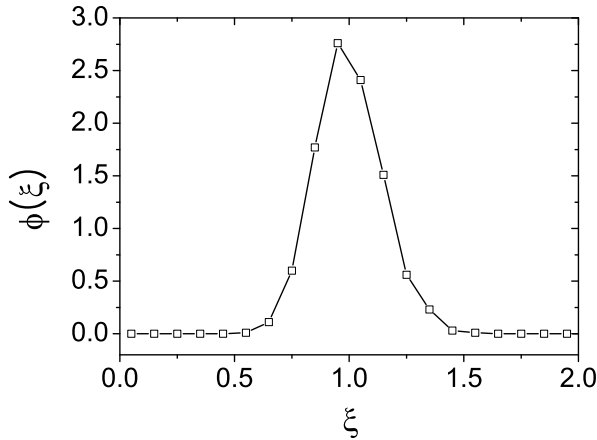}
\caption{Scatter distribution $\phi_{\alpha}(\xi)$ for CTRW subdiffusion with
cutoff, as defined in Eq.~(\ref{wtd_cutoff}). The observed form is
clearly different from the much more asymmetric shape in
Fig.~\ref{ctrw_scatter}.
The value of $\phi_{\alpha}(\xi)$ reaches zero
for small values of $\xi$. An almost Gaussian shape centred around the
ergodic value $\xi=1$ is observed. Same parameters as in
Fig.~\ref{ctrw_cutoff_conf}.}
\label{cutoff_scatter}
\end{figure}

The scatter distribution is sensitive to few extreme events. When these are
lacking, the distribution should be significantly different from the form
discussed for CTRW subdiffusion with diverging characteristic time scale.
Indeed, as shown in Fig.~\ref{cutoff_scatter} for the cutoff waiting time
distribution defined through Eq.~(\ref{wtd_cutoff}), $\phi_{\alpha}(\xi)$
drops down to zero around $\xi=0$ and assumes an almost Gaussian shape
around the ergodic value $\xi=1$.
Note that the simulations were performed with the same parameters as for
Fig.~\ref{ctrw_cutoff_conf}, in which the breaking of ergodicity still
persists, despite the cutoff. This demonstrates that different quantities
have different sensitivity to the cutoff.

\subsection{Fractional Brownian motion}

\begin{figure}
\includegraphics[width=8.8cm]{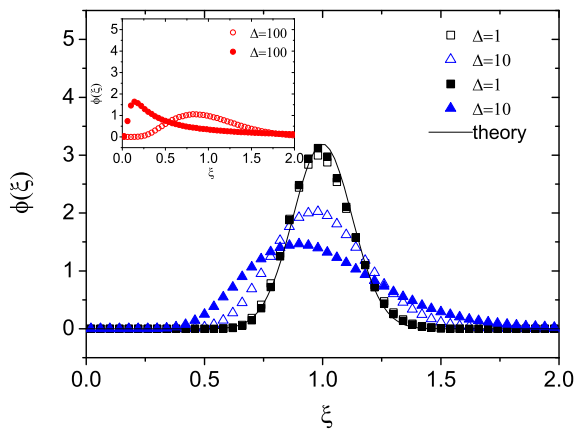}
\caption{Scatter distribution $\phi(\xi)$ of the time averaged mean
squared displacement, with $\xi=\overline{\delta^2}/\left<\overline{\delta^2}
\right>$ for FBM subdiffusion with $\alpha=0.5$ and $T=128$ (a.u.). The filled and
open symbols, respectively, represent the cases for unconfined and confined
motion. For small values of the lag time $\Delta$ we see quite good agreement
with the expected Gaussian limiting distribution of the scatter,
Eq.~(\ref{fbmscatter}), that is centred around $\xi=1$. For larger values
of $\Delta$ deviations are observed, however, even for $\Delta=100$ the value
around $\xi=0$ is consistently zero. A similar behaviour is found for larger
values of $\alpha$ \cite{jae_scatter}. Note that when $T\to\infty$ the
process is ergodic and $\phi(\xi)$ approaches a $\delta$ function.}
\label{fbm_scatter}
\end{figure}

FBM is based on long-ranged correlations. However, we can obtain an
approximate expression for the scatter distribution \cite{jae_scatter}
\begin{equation}
\label{fbmscatter}
\phi(\xi)\approx\sqrt{\frac{T-\Delta}{4\pi\tau^{\dagger}}}\exp\left(
-\frac{(\xi-1)^2(T-\Delta)}{4\tau^{\dagger}}\right).
\end{equation}
where $\tau^{\dagger}$ is an intrinsic time scale. Expression
(\ref{fbmscatter}) is valid for sufficiently small lag times $\Delta$,
for which the correlations are neglected \cite{jae_scatter}. Note
that this distribution is independent of $\alpha$ and centred around the
ergodic value $\xi=1$. In particular, in the long measurement time limit
$T\to\infty$, the distribution converges to the sharp form $\phi(\xi)=
\delta(\xi-1)$. Also this behaviour is surprisingly well preserved for short
trajectories, as demonstrated in Fig.~\ref{fbm_scatter}.

For FBM one can define a quantity similar to the ergodicity breaking parameter
(\ref{eb}). Namely, without taking the long measurement time limit, we obtain
the normalised variance of the time averaged mean squared displacement
\begin{equation}
\label{v}
V=\frac{\left<\left(\overline{\delta^2(\Delta,T)}\right)^2\right>-\left<
\overline{\delta^2(\Delta,T)}\right>^2}{\left<\overline{\delta^2(\Delta,T)}
\right>^2}.
\end{equation}
It turns out that the convergence to ergodicity is algebraically slow, and for
subdiffusion $0<\alpha<1$ one obtains a decay, which is inversely proportional
to $T$ \cite{deng}:
\begin{equation}
V\sim k(\alpha)\frac{\Delta}{T}.
\end{equation}
The coefficient $k(\alpha)$ here is defined as \cite{deng}
\begin{equation}
k(\alpha)=\int_0^{\infty}\Big((t+1)^{\alpha}+|t-1|^{\alpha}-2t^{\alpha}\Big)dt.
\end{equation}
It increases continuously from zero at $\alpha=0$ to $k(1)=4/3$.

The quantity $V$ is connected to the probability density (\ref{fbmscatter})
by $V=2\tau^{\dagger}/(T-\Delta)\sim2\tau^{\dagger}/T$, obtained from
approximating
that the values of $\xi$ may range in the interval $(-\infty,\infty)$, which
appears reasonable given the sharp decay of $\phi(\xi)$ to 0 at $\xi=0$ for
FBM. Apart from the fact that $V$ explicitly depends on $\alpha$ while the
approximation leading to Eq.~(\ref{fbmscatter}) loses the $\alpha$ dependence,
both quantities decay $\sim1/T$, and the internal time scale $\tau^{\dagger}$
takes on a role similar to the lag time $\Delta$.

\section{Velocity autocorrelation functions}

A typical quantity accessible from experimental data is the velocity
autocorrelation function, which is defined through
\begin{equation}
 C_{v}^{(\epsilon)}(\tau)=\frac{1}{\epsilon^2} \Big<
\left(x(\tau+\epsilon)-x(\tau)\right)\left(x(\epsilon)-x(0)\right)\Big>.
\label{autov00}
\end{equation}
Here the velocity is defined as the difference quotient $v(\tau)=\epsilon^
{-1}[x(\tau+\epsilon)-x(\tau)]$. The velocity autocorrelation (\ref{autov00})
was suggested as a tool to distinguish between different subdiffusion
models \cite{weber}.
We show here that for confined processes the shape
of the velocity autocorrelation function does not allow for a significant
distinction between subdiffusive CTRW and FBM. Note that the calculation
of $C_v^{(\epsilon)}(\tau)$ amounts to determine four two-point correlation
functions of the type $\langle x(t_1)x(t_2)\rangle$.

\subsection{Continuous time random walk}

\begin{figure}
\includegraphics[height=8.8cm,angle=-90]{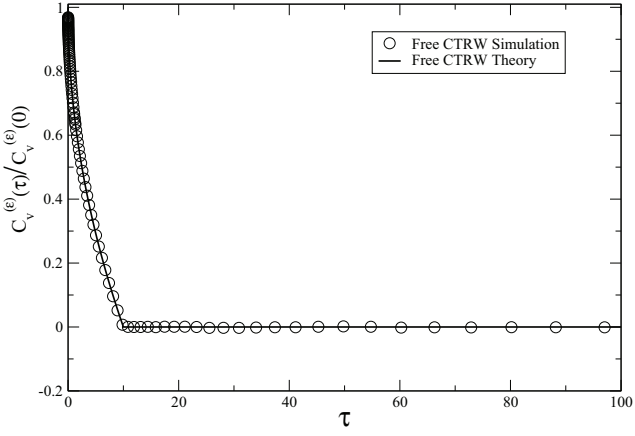}
\caption{Normalised velocity autocorrelation function $C_{v}^{(\epsilon)}
(\tau)$ for an unconfined CTRW process with $\alpha=0.5$. The simulations
($\bigcirc$) were performed for time 10000 (a.u) with $\epsilon=10$ (a.u).
The theoretical behaviour (---) is given by Eq.~(\ref{autov04}).}
\label{velcorr1}
\end{figure}

For unbounded CTRW process starting with initial condition $x(0)=0$, the
position correlation function is given by the following expression
\cite{Friedrich},
\begin{equation}
\langle x(t_1)x(t_2)\rangle=\frac{2K_\alpha}{\Gamma(1+\alpha)}\Big[\min\{t_1,
t_2\}\Big]^\alpha.
\label{autov03}
\end{equation}
This result for free CTRW is due to the fact that the jump lengths in the
interval $(t_1,t_2)$ for $t_2>t_1$ have zero mean. With Eq.~(\ref{autov03})
we find
\begin{equation}
\frac{C_{v}^{(\epsilon)}(\tau)}{C_{v}^{(\epsilon)}(0)}=\left\{
\begin{array}{ll}\displaystyle\frac{\epsilon^\alpha-\tau^\alpha}{\epsilon^
\alpha} & \tau\leq\epsilon \\[0.32cm]
0 & \tau\geq\epsilon \end{array}\right.
\label{autov04}
\end{equation}
The velocity autocorrelation function for an unbounded CTRW process does not
yield negative values due to the absence of correlations for different jumps.
In this case the velocity autocorrelations can be easily distinguished from
the behaviour of FBM processes, see below. Note that $C_v^{(\epsilon)}(\tau)$
in Eq.~(\ref{autov04}) is non-analytic at $\tau=\epsilon$, an observation that
is confirmed in simulations. The behaviour of the velocity autocorrelations for
CTRW subdiffusion are displayed in Fig.~\ref{velcorr1}.

\begin{figure}
\includegraphics[height=8.8cm,angle=-90]{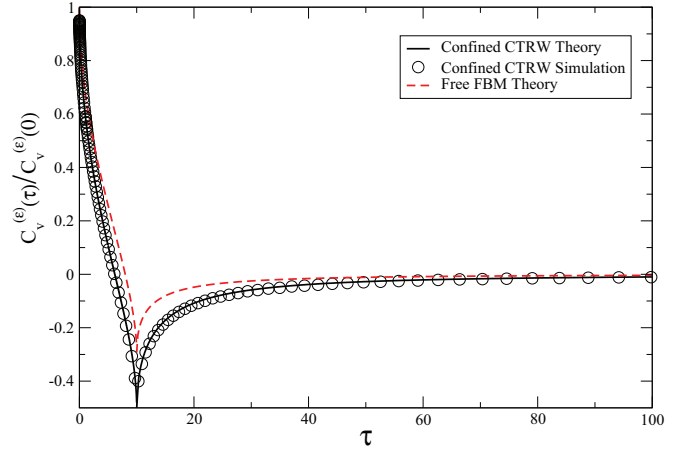}
\caption{Normalised velocity autocorrelation function $C_{v}^{(\epsilon)}
(\tau)$ for a free FBM process and a confined CTRW process with $\alpha=0.5$.
The CTRW simulations ($\bigcirc$) were performed over the time range 10000
(a.u), and the system corresponds to a particle on a lattice with ten
lattice points and reflecting boundaries. We chose $\epsilon=10$ (a.u).
The theoretical prediction for the CTRW (---) is given by Eq.~(\ref{autov05}),
and for FBM (- -) by Eq.~(\ref{autov01}).}
\label{velcor2}
\end{figure}

\begin{figure}
\includegraphics[height=8.8cm,angle=-90]{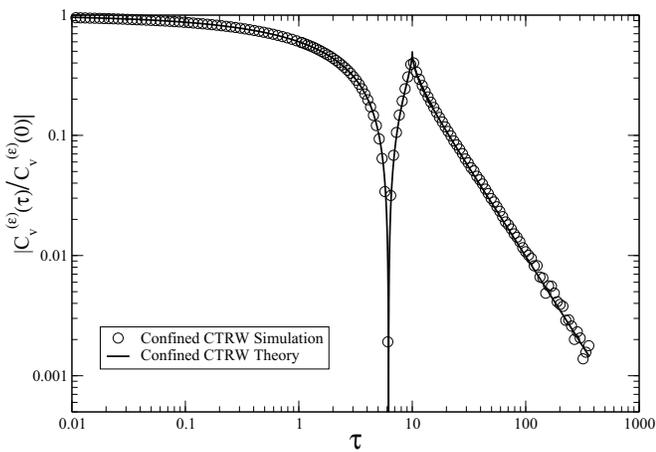}
\caption{Absolute value of the normalised velocity autocorrelation function
$C_{v}^{(\epsilon)}(\tau)$ for confined CTRW process with $\alpha=0.5$. The
CTRW simulations ($\bigcirc$) are based on the parameters from
Fig.~\ref{velcor2}. The theory line (---) corresponds to Eq.~(\ref{autov05}).}
\label{velcor3}
\end{figure}

For a confined CTRW process the situation is quite different. To explore the
behaviour of the corresponding velocity autocorrelation function we use the
general result for the correlation function of confined CTRW,
Eq.~(\ref{ctrw_conf_ea_auto}). For simplicity we assume a symmetric potential
such that $\langle x\rangle_B=0$. With the initial condition $x(0)=0$ we
obtain
\begin{eqnarray}
\nonumber
\hspace*{-0.8cm}
\frac{C_{v}^{(\epsilon)}(\tau)}{C_{v}^{(\epsilon)}(0)}&=&
\frac{1}{\Gamma(\alpha)\Gamma(1-\alpha)}\\
&&\hspace*{-1.6cm}
\times\left\{
\begin{array}{ll}
B\left(\frac{\epsilon}{\epsilon+\tau},\alpha,1-\alpha\right)
-B\left(\frac{\tau}{\epsilon},\alpha,1-\alpha\right) & \epsilon\geq\tau \\
B\left(\frac{\epsilon}{\epsilon+\tau},\alpha,1-\alpha\right)-B\left(\frac{
\epsilon}{\tau},\alpha,1-\alpha\right) & \epsilon\leq\tau \end{array}\right..
\label{autov05}
\end{eqnarray}
Figs.~\ref{velcor2} and \ref{velcor3} (for the absolute value) display
excellent agreement of 
Eq.~(\ref{autov05}) with simulations of CTRW subdiffusion for a lattice of
size 10 with reflecting boundary conditions, and $\alpha=1/2$. We observe
that for confined motion the CTRW velocity autocorrelation function indeed
attains negative values and has a minimum on $\tau=\epsilon$, as the
confinement effectively induces correlations. For long $\tau$
the velocity autocorrelation function decays to zero (from the negative side)
as the power-law 
\begin{equation}
\frac{C_{v}^{(\epsilon)}(\tau)}{C_{v}^{(\epsilon)}(0)}\sim  
-\frac{1}{\Gamma(\alpha)\Gamma(1-\alpha)}\left(\frac{\epsilon}{\tau}\right)^{
1+\alpha},
 \label{autov06}
\end{equation}
valid for $\tau\gg\epsilon$.

\subsection{Fractional Brownian Motion}

For free FBM the position correlation function (\ref{fbm_auto}) together with
the definition (\ref{autov00}) produce the result
\begin{equation}
\frac{C_{v}^{(\epsilon)}(\tau)}{C_{v}^{(\epsilon)}(0)}=\frac{(\tau+\epsilon)^
\alpha-2\tau^\alpha+|\tau-\epsilon|^\alpha}{2\epsilon^\alpha}.
\label{autov01}
\end{equation}
This function yields negative values for sufficiently long $\tau$, and its
minimum value $(2^{\alpha-1}-1)$ is assumed at $\tau=\epsilon$.  For
long $\tau$ it decays toward zero from the negative side in the power-law
form
\begin{equation}
\frac{C_{v}^{(\epsilon)}(\tau)}{C_{v}^{(\epsilon)}(0)}\sim-\frac{\alpha-
\alpha^2}{2} \left(\frac{\epsilon}{\tau}\right)^{2-\alpha}
\label{autov02}
\end{equation}
valid for $\tau\gg\epsilon$.

\begin{figure}
\includegraphics[width=8.8cm]{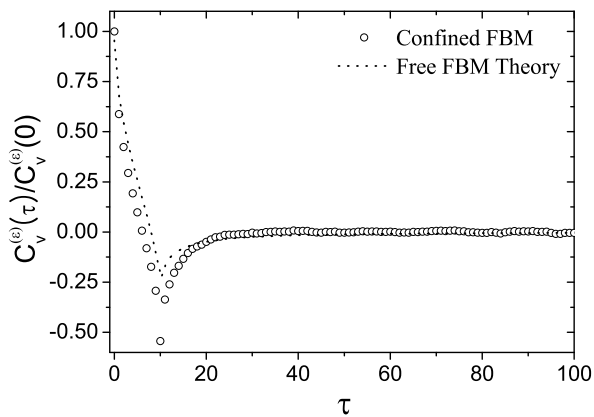}
\caption{Normalised velocity autocorrelation function $C_{v}^{(\epsilon)}
(\tau)$ for confined FBM with $\alpha=0.5$. The simulations were performed
on the interval $[-2,2]$, and $\epsilon=10$ (a.u.). For comparison, the
velocity autocorrelation for free FBM is drawn based on Eq.~(\ref{autov01}).}
\label{fbm_box}
\end{figure}

We see that both the velocity autocorrelation
function of the confined CTRW and the one for free FBM acquire negative
values and decay as power-laws for large $\tau$, see below. Further both
obtain a sharp minimum for $\tau=\epsilon$. Confined FBM behaves similarly
to free FBM, see Fig.~\ref{fbm_box}. From our analysis it becomes
clear that the shape of the velocity autocorrelation function may not be a
good diagnosis tool to distinguish between subdiffusive CTRW and FBM
processes. We note that for CTRW subdiffusion the time averaged correlation
functions will be random variables, exactly like the time averaged mean
squared displacement.

\section{Discussion}

The physical mechanisms leading to subdiffusion in biological cells or other
complex systems are varied. The anomalous diffusion of inert biopolymers larger
than some 10kD in living cells is due to molecular crowding, these are excluded
volume effects in the viscoelastic, superdense cellular environment
\cite{minton,zimmermann,mcguffee}. Potentially similar effects occur in
single file diffusion: A tracer particle diffusing in a one dimensional
channel and interacting with other Brownian particles will slow down its
random motion, and subdiffusion with $\alpha=1/2$ is observed \cite{singlefile}.
What is the resulting motion like? Consider a particle, that gets repeatedly
trapped during its random walk. Such trapping may give rise to a broad
distribution of waiting times, as embodied in the CTRW model. Such broad
distribution of trapping times could be due to chemical binding of the tracer
particle to its environment on varied time scales, or due to active
gelation/de-gelation of the crowding particles. Long-tailed trapping time
distributions were observed for $\mu$m-sized particles due to interaction
with a semiflexible actin mesh \cite{weitz}. There, the anomalous diffusion
exponent $\alpha$ depends on the size of the tracer particles versus the
typical actin mesh size. Alternatively, motion patterns of FBM type may be
due to coupling to the viscoelastic crowding environment \cite{weiss1}.
Yet other approaches are based on polymer dynamics. De Gennes' reptation
model of a polymer diffusing in a tube formed by foreign chains in a melt
yields subdiffusion with $\alpha=1/4$, and was used as a possible explanation
of the short time dynamics of telomere motion \cite{garini}.

One may speculate whether the widely observed subdiffusion of biomolecules and
passive tracers is a mere coincidence of nature that should be attributed to
the dense environment in the cell and/or to the wide distribution of obstacles
and reactions in the cell. Or, maybe, subdiffusion is by itself a goal which
was obtained via evolution? There exist claims that sub-diffusion is useful
in certain cellular search strategies \cite{golding,guigas}. One may also view
subdiffusion to be a sufficiently slow process to help maintaining the
organisation of the genome in the nucleus of the cell without the need for
physical compartments \cite{garini}.
Since reactions in many cases are controlled by diffusion, the emergence of
slower-than-normal diffusion has far-reaching implications to signalling and
regulatory processes in the cell. The standard theories of diffusion-controlled
reactions under such circumstances must be replaced by subdiffusion-controlled
reaction models. While these issues are clearly important, we here focused
more on the characterisation of trajectories of single particles in the cell,
in particular, as a quantitative diagnosis method to probe the nature of the
stochastic motion in cells and other complex media.

Single particle tracking microscopy is a widely used technique. It allows
one to locally probe complex systems in the liquid phase \emph{in situ}. In
particular it has become one of the standard tools in biophysical, colloidal,
polymeric, and gel-like environments. The motion of tracers in these
systems is often anomalous. Typically the experimentally recorded time series
are evaluated in terms of the time averaged mean squared displacement
$\overline{\delta^2(\Delta,T)}$. Here we collected the behaviour of $\overline{
\delta^2(\Delta,T)}$ for the two most prominent anomalous stochastic processes,
the continuous time random walk and fractional Brownian motion.

For CTRW subdiffusion, connected with ageing and weak ergodicity breaking,
in an unconfined system the time averaged mean squared displacement scales
linearly in time, $\overline{\delta^2(\Delta,T)}$ [Eq.~(\ref{ctrwtamsd})],
renouncing the anomalous nature of the process. Only the dependence on the
overall measurement time $T$ pays tribute to the underlying subdiffusion.
This behaviour contrasts the anomalous scaling $\langle x^2(t)\rangle=2K_
{\alpha}t^{\alpha}$ of the ensemble averaged mean squared displacement.
Under confinement no plateau value is observed as in the ensemble average,
instead, a power-law of the
form $\overline{\delta^2(\Delta,T)}\simeq\Delta^{1-\alpha}$,
Eq.~(\ref{conf_tamsd}), is found. Interestingly, this characteristic behaviour
is approximately preserved when an appropriate cutoff in the waiting time is
introduced. Conversely the distribution of the time averaged mean squared
displacement around its ensemble mean becomes almost Gaussian in presence of
the cutoff, while it has an exponential decay and a finite value at $\xi=0$
when the system is non-ergodic and exhibits ageing.

FBM, in contrast, is ergodic, although ergodicity is reached algebraically
slowly. Free FBM subdiffusion shows $\overline{\delta^2(\Delta,T)}\simeq\Delta
^{\alpha}$, Eq.~(\ref{fbm_free}), while under confinement the plateau value
of the ensemble average is reached, Eq.~(\ref{fbm_cnf}).
For finite trajectories the distribution of the time averaged mean
squared displacement is approximately Gaussian for short lag times.

Our analysis demonstrates how different the two stochastic processes are,
despite sharing the same form of the ensemble averaged mean squared
displacement. At the same time the velocity autocorrelation of confined
CTRW subdiffusion is hardly distinguishable from that of FBM subdiffusion.
Given a recorded time series of anomalous diffusion from
experiment it is important to know more precisely which stochastic process
is responsible for the observed behaviour, in particular, with respect to
diffusion-limited reactions, general transport behaviour, and related
processes such as gene regulation. The analysis presented here, along with
complementary tools discussed in Refs.~\cite{christine,vincent,marcin}, will
be instrumental in the classification of anomalous diffusion behaviour.

\acknowledgments

We thank C. Selhuber-Unkel and L. Oddershede for providing the data used in
Fig.~\ref{christine}, and for many useful discussions.
Partial financial support from the Deutsche Forschungsgemeinschaft and
the Israel Science Foundation is gratefully acknowledged.

\begin{appendix}

\section{Scatter distribution}
\label{app1}

Using the theory of Fox $H$-functions we obtain the exact form for the
distribution $\phi_{\alpha}(\xi)$ of the dimensionless variable $\xi$
\cite{mathai}:
\begin{equation}
\phi_{\alpha}(\xi)=\frac{1}{\alpha^2\xi}H^{1,0}_{1,1}\left[\frac{\xi^{1/\alpha}}
{\Gamma(1+\alpha)^{1/\alpha}}\left|\begin{array}{l}(0,1)\\(0,1/\alpha)
\end{array}\right.\right].
\end{equation}
This function has the series expansion
\begin{equation}
\phi_{\alpha}(\xi)=\frac{1}{\alpha\xi}\sum_{n=0}^{\infty}\frac{(-1)^n[\Gamma(1+
\alpha)/\xi]^n}{n!\Gamma(-\alpha n)},
\end{equation}
and the asymptotic form
\begin{eqnarray}
\nonumber
\phi_{\alpha}(\xi)&\simeq&\left(\xi^{1/\alpha}\right)^{(1-\alpha)/(2\alpha)-
\alpha}\\
&&\hspace*{-0.2cm}
\times\exp\left(-(1-\alpha)\left[\frac{\alpha^{\alpha}\xi}{\Gamma(1+\alpha)}
\right]^{1/(1-\alpha)}\right)
\end{eqnarray}
valid at $\xi\gg\Gamma(1+\alpha)$.




\end{appendix}


\begin{thebibliography}{99}

\bibitem{braeuchle} C. Br{\"a}uchle, D. C. Lamb, and J. Michaelis, Editors,
Single Particle Tracking and Single Molecule Energy Transfer (Wiley-VCH,
Weinheim, 2010).

\bibitem{saxton} M. J. Saxton and K. Jacobson, Ann. Rev. Biophys. Biomol.
Struct. \textbf{26}, 373 (1997).

\bibitem{saxton1} M. J. Saxton, Biophys. J. \textbf{72}, 1744 (1997).

\bibitem{qian} H. Qian, M. P. Sheetz, and E. L. Elson, Biophys. J. \textbf{60},
910 (1991).

\bibitem{perrin} J. Perrin, Comptes Rendus (Paris) \textbf{146}, 967 (1908);
Ann. Chim. Phys. \textbf{18}, 5 (1909).

\bibitem{nordlund} I. Nordlund, Z. Physik. Chemie \textbf{87}, 40 (1914).

\bibitem{microrheol} T. G. Mason and D. A. Weitz, Phys. Rev. Lett. \textbf{74},
1250 (1995).

\bibitem{motors} W. J. Greenleaf, M. T. Woodside, and S. M. Block, Ann. Rev.
Biophys. Biomol. Struct. \textbf{36}, 171 (2007).

\bibitem{golding} I. Golding and E. C. Cox, Phys. Rev. Lett. {\bf 99}, 098102
(2006).

\bibitem{weber} S. C. Weber, A. J. Spakowitz, and J. A. Theriot, Phys. Rev.
Lett. \textbf{104}, 238102 (2010).

\bibitem{garini} I. Bronstein, Y. Israel, E. Kepten, S. Mai, Y. Shav-Tal,
E. Barkai, and Y. Garini, Phys. Rev. Lett. \textbf{103},018102 (2009).

\bibitem{seisenhuber} G. Seisenberger, M. U. Ried, T. Endre{\ss}, H.
B{\"u}ning, M. Hallek, C. and Br{\"a}uchle, Science \textbf{294}, 1929
(2001).

\bibitem{elbaum} A. Caspi, R. Granek, and M. Elbaum, Phys. Rev. Lett.
\textbf{85}, 5655 (2000).

\bibitem{lene} I. M. Toli{\'c}-N{\o}rrelykke, E. L. Munteanu, G. Thon, L.
Oddershede, and K. Berg-S{\o}rensen, Phys. Rev. Lett. {\bf 93}, 078102 (2004).

\bibitem{lene1} C. Selhuber-Unkel, P. Yde, K. Berg-S{\o}rensen, and
L. B. Oddershede, Phys. Biol. \textbf{6}, 025015 (2009).

\bibitem{weihs} N. Gal and D. Weihs, Phys. Rev. E \textbf{81}, 020903(R) (2010).

\bibitem{banks} D. Banks and C. Fradin, Biophys. J. \textbf{89}, 2960 (2005).

\bibitem{weiss} M. Weiss, M. Elsner, F. Kartberg, and T. Nilsson, Biophys. J.
\textbf{87}, 3518 (2004).

\bibitem{weiss1} J. Szymanski and M. Weiss, Phys. Rev. Lett. \textbf{103},
038102 (2009).

\bibitem{yves} J. Vercammen, G. Martens, and Y. Engelborghs, Springer
Ser. Fluoeresc. \textbf{4}, 323 (2007).

\bibitem{pan} W. Pan, L. Filobelo, N. D. Q. Pham, O. Galkin, V. V. Uzunova,
and P. G. Vekilov, Phys. Rev. Lett. \textbf{102}, 058101 (2009).

\bibitem{weitz} I. Y. Wong, M. L. Gardel, D. R. Reichman, E. R. Weeks, M. T.
Valentine, A. R. Bausch, and D. A. Weitz, Phys. Rev. Lett. \textbf{92}, 178101
(2004).

\bibitem{christine} J.-H. Jeon, V. Tejedor, S. Burov, E. Barkai, C. Selhuber,
K. Berg-S{\o}rensen, L. Oddershede, and R. Metzler (unpublished).

\bibitem{vankampen} N. G. van Kampen, Stochastic processes in physics and
chemistry (Elsevier, Amsterdam, 2007).

\bibitem{khinchin} A. Y. Khinchin, Mathematical foundations of statistical
mechanics (Dover Publications Inc., New York, NY, 2003).

\bibitem{sinai} Y. Sinai, Theor. Prob. Appl. \textbf{27}, 256 (1982);
J. Dr{\"a}ger and J. Klafter, Phys. Rev. Lett. \textbf{84}, 5998 (2000).

\bibitem{tejedor}  V. Tejedor and R. Metzler, J. Phys. A \textbf{43},
082002 (2010).

\bibitem{goychuk} P. Siegle, I. Goychuk, and P. H{\"a}nggi, Phys. Rev. Lett.
\textbf{105}, 100602 (2010).

\bibitem{report} R. Metzler and J. Klafter, Phys. Rep. \textbf{339}, 1 (2000);
J. Phys. A \textbf{37}, R161 (2004); E. Barkai, Phys. Rev. E \textbf{63},
046118 (2001).

\bibitem{zimmermann} S. B. Zimmerman and S. O. Trach, J. Mol. Biol.
\textbf{222}, 599 (1991).

\bibitem{minton} R. J. Ellis and A. P. Minton, Nature \textbf{425}, 27
(2003); A. P. Minton, J. Cell Science \textbf{199}, 2863 (2006).

\bibitem{mcguffee} S. R. McGuffee and A. H. Elcock, PLoS Comput. Biol.
\textbf{6}, e1000694 (2010).

\bibitem{katja} S. B. Yuste, G. Oshanin, K. Lindenberg, O. B{\'e}nichou, and
J. Klafter, Phys. Rev. E {\bf 78}, 021105 (2008); E. Abad. S. B. Yuste, and
K. Lindenberg, Phys. Rev. E {\bf 81}, 031115 (2010); I. M. Sokolov, S. B. Yuste,
J. J. Ruiz-Lorenzo, and K. Lindenberg, Phys. Rev. E {\bf 79}, 051113 (2009);
D. Froemberg and I. M. Sokolov, Phys. Rev. Lett. \textbf{100}, 108304 (2008).

\bibitem{he} Y. He, S. Burov, R. Metzler, and E. Barkai, Phys. Rev. Lett.
\textbf{101}, 058101 (2008).

\bibitem{pnas} S. Burov, R. Metzler, and E. Barkai, Proc. Natl. Acad. Sci.
USA \textbf{107}, 13228 (2010).

\bibitem{appb} R. Metzler, V. Tejedor, J.-H. Jeon, Y. He, W. Deng, S. Burov,
and E. Barkai, Acta Phys. Polonica B \textbf{40}, 1315 (2009).

\bibitem{havlin} S. Havlin and D. ben-Avraham, Adv. Phys. {\bf 36},
695 (1987).

\bibitem{kimmich} A. Klemm, R. Metzler, and R. Kimmich, Phys. Rev. E
\textbf{65}, 021112 (2002); see also References therein.

\bibitem{montroll} E. W. Montroll and G. H. Weiss, J. Math.
Phys. {\bf 10}, 753 (1969).

\bibitem{scher} H. Scher and E. W. Montroll, Phys. Rev. B \textbf{12}, 2455
(1975).

\bibitem{scher1} H. Scher, G. Margolin, R. Metzler, J. Klafter, and B.
Berkowitz, Geophys. Res. Lett. \textbf{29}, 1061 (2002);
B. Berkowitz, A. Cortis, M. Dentz and H. Scher, Reviews of Geophysics, 44,
RG2003 (2006).

\bibitem{rel} C. Monthus and J.-P. Bouchaud, J. Phys. A \textbf{29}, 3847
(1996); G. Ben Arous, A. Bovier, and V. Gayrard Phys. Rev. Lett. \textbf{88},
087201 (2002)

\bibitem{ageing} E. Barkai and Y. C. Cheng, J. Chem. Phys. \textbf{118},
6167 (2003).

\bibitem{web} J.-P. Bouchaud, J. Phys. (Paris) I, \textbf{2}, 1705 (1992);
G. Bel and E. Barkai, Phys. Rev. Lett. \textbf{94}, 240602 (2005); A.
Rebenshtok and E. Barkai, \emph{ibid.} \textbf{99}, 210601 (2007); M. A. 
Lomholt, I. M. Zaid, and R. Metzler, \emph{ibid.}, 98, 200603 (2007);
I. M. Zaid, M. A. Lomholt, and R. Metzler, Biophys. J. \textbf{97}, 710 (2009);
F. D. Stefani, J. P. Hoogenboom, and E. Barkai, Physics Today \textbf{62}(2),
34 (2009); G. Margolin and E. Barkai, Phys. Rev. Lett. \textbf{94}, 080601
(2005); G. Aquino, P. Grigolini, and B. J. West, EPL \textbf{80}, 10002 (2007).

\bibitem{mebakla} R. Metzler, E. Barkai, and J. Klafter, Phys. Rev. Lett.
\textbf{82}, 3563 (1999); Europhys. Lett. \textbf{46}, 431 (1999).
E. Barkai, R. Metzler, and J. Klafter, Phys. Rev. E
\textbf{61}, 132 (2000); R. Metzler, J. Klafter, and I. M. Sokolov,
\emph{ibid.} \textbf{58}, 1621 (1998).


\bibitem{rel1} S. Burov and E. Barkai, Phys. Rev. Lett. \textbf{98}, 250601
(2007).

\bibitem{hughes} B. D. Hughes {\it Random Walks and Random Environments,
Volume 1: Random Walks\/} (Oxford University Press, Oxford, 1995).

\bibitem{bouchaud} J.-P. Bouchaud and A. Georges, Phys. Rep. \textbf{195},
127 (1990).

\bibitem{scher2} H. Scher, M. F. Shlesinger, and J. T. Bendler, Phys. Today
\textbf{44}, No.~1, 26 (1991).

\bibitem{klages} Anomalous transport: foundations and applications, edited by
R. Klages, G. Radons, and I. M. Sokolov (Wiley-VCH, Weinheim, 2007).

\bibitem{fbm} A. N. Kolmogorov, Dokl. Acad. Sci. USSR \textbf{26},
115 (1940); B. B. Mandelbrot and J. W. van Ness, SIAM Rev. \textbf{1},
422 (1968); H. Qian, Fractional Brownian Motion and Fractional
Gaussian Noise. In G. Rangarajan and M.Z. Ding (eds), {\it Processes
with Long-Range Correlations} (Springer, Lecture Notes in Physics,
Vol.621), pp.22-33.

\bibitem{kubo} R. Kubo, M. Toda, and N. Hashitsume, Statistical Physics II:
Nonequilibrium Statistical Mechanics (Springer-Verlag, Heidelberg, 1995).

\bibitem{hurst} H. E. Hurst, Trans. Am. Soc. Civ. Eng. \textbf{116}, 400 (1951).

\bibitem{climate} T. N. Palmer, G. J. Shutts, R. Hagedorn, F. J. Doblas-Reyes,
T. Jung, and M. Leutbecher, Annu. Rev. Earth Planet Sci. \textbf{33}, 163
(2005).

\bibitem{eco} I. Simonsen, Physica A \textbf{322}, 597 (2003); N. E. Frangos,
S. D. Vrontos, and A. N. Yannacopoulos, Appl. Stoch., Models Bus. Ind.
\textbf{23}, 403 (2007).

\bibitem{singlefile} T. E. Harris, J. Appl. Probab. \textbf{2}, 323 (1965);
L. Lizana and T. Ambj{\"o}rnsson, Phys. Rev. Lett. \textbf{100},
200601 (2008); L. Lizana, T. Ambj{\"o}rnsson, A. Taloni, E. Barkai, and
M. A. Lomholt, Phys. Rev. E \textbf{81}, 051118 (2010); E. Barkai and R.
Silbey, Phys. Rev. Lett. \textbf{102}, 050602 (2009).

\bibitem{chechkin} A. Taloni, A. V. Chechkin, and J. Klafter, Phys. Rev. Lett.
\textbf{104}, 160602 (2010).

\bibitem{xie} S. C. Kou and X. S. Xie, Phys. Rev. Lett. \textbf{93}, 180603
(2004); W. Min, G. Luo, B. J. Cherayil1, S. C. Kou, and X. S. Xie,
Phys. Rev. Lett. \textbf{94}, 198302 (2005).

\bibitem{marcin} M. Magdziarz, A. Weron, K. Burnecki, and J. Klafter, Phys.
Rev. Lett. \textbf{103}, 180602 (2009). Compare also M. Magdziarz and J.
Klafter, Phys. Rev. E, at press; K. Burnecki and A. Weron, Phys. Rev. E
\textbf{82}, 021130 (2010).

\bibitem{lubelski} A. Lubelski, I. M. Sokolov, and J. Klafter, Phys. Rev. Lett.
\textbf{100}, 250602 (2008).


\bibitem{deng} W. H. Deng and E. Barkai, Phys. Rev. E \textbf{79}, 011112
(2009).

\bibitem{subord} H. C. Fogedby, Phys. Rev. E \textbf{50}, 1657 (1994); A.
Baule and R. Friedrich, Phys. Rev. E \textbf{71}, 026101 (2005);
M. Magdziarz, A. Weron, and J. Klafter, Phys. Rev. Lett. \textbf{101},
210601 (2008); M. Magdziarz, A. Weron, and K. Weron, Phys. Rev. E \textbf{75},
016708 (2007); E. Heinsalu, M. Patriarca, I. Goychuk, G. Schmid, and P.
H{\"a}nggi, Phys. Rev. E \textbf{73}, 046133 (2006).

\bibitem{igor} T. Neusius, I. M. Sokolov, and J. C. Smith, Phys. Rev. E
\textbf{80}, 011109 (2009).

\bibitem{jae} J.-H. Jeon and R. Metzler, Phys. Rev. E \textbf{81}, 021103
(2009).

\bibitem{oleksii} O. Y. Sliusarenko, V. Y. Gonchar, A. V. Chechkin, I. M.
Sokolov, and R. Metzler, Phys. Rev. E \textbf{81}, 041119 (2010).

\bibitem{igor_epl} I. M. Sokolov, E. Heinsalu, P. H{\"a}nggi, and I.
Goychuk, EPL \textbf{86}, 30009 (2009).

\bibitem{jae_scatter} J.-H. Jeon and R. Metzler, J. Phys. A \textbf{43}, 252001
(2010).

\bibitem{Friedrich} A. Baule and R. Friedrich  {Eur. Phys. Lett.} {\bf 77},
10002 (2007).

\bibitem{guigas} G. Guigas and M. Weiss, Biophys. J. \textbf{94}, 90 (2008).

\bibitem{vincent} V. Tejedor, O. B{\'e}nichou, R. Voituriez, R. Jungmann, F.
Simmel, C. Selhuber, L. Oddershede, and R. Metzler, Biophys. J. \textbf{98},
1364 (2010).

\bibitem{mathai} A. M. Mathai and R. K. Saxena, The H-function
with Applications in Statistics and Other Disciplines (Wiley Eastern
Ltd., New Delhi, 1978); A. M. Mathai, R. K. Saxena, and H. J. Haubold,
The H-function, Theory and Applications (Springer, New York, 2010).

\end{thebibliography}
\end{document}